\documentstyle[aps,preprint,epsf]{revtex}
\begin{document}
\tighten
\title{Semianalytic Analysis of Primordial Black Hole Formation
During a First-order QCD Phase Transition}
\draft
\author{Christian Y. Cardall}
\address{Department of Physics \& Astronomy,
State University of New York at Stony Brook,
Stony Brook, NY 11794-3800}
\author{George M. Fuller}
\address{Department of Physics, University of California, San
Diego, La Jolla, CA 92093-0319}
\date{\today}
\maketitle
\begin{abstract}
It has recently been suggested that cosmologically significant
numbers of black holes could form  during a first-order QCD phase  
transition.
Further, it has been asserted that  these
black holes would have masses
corresponding naturally to the inferred mass
($\sim 1 M_{\odot}$) of the MACHOs
responsible for the observed gravitational microlensing events.
In this model, the underlying spectrum of primordial 
density perturbations
provides the fluctuations that give rise to black holes at the  
epoch of the QCD transition.
We employ a simplified model to estimate the reduction in the
critical overdensity of  a horizon-sized primordial
perturbation required for collapse to
a black hole. We find that a first-order QCD transition does
indeed produce a sharp peak in the black hole mass
spectrum, but that this peak corresponds to the horizon mass
at an epoch somewhat earlier than the cosmological transition
itself.  Assuming a COBE normalized primordial density 
perturbation spectrum with constant spectral index, 
for the black holes so produced to be cosmologically 
significant would require an extremely finely tuned ``blue'' 
primordial density perturbation spectrum. Specifically,
in the context of our simplified model,
a spectral index in the range $n=1.37-1.42$ corresponds to
the range $\Omega \sim 10^{-5}-10^3$ of the 
black hole contribution
to the present-day density parameter.
\end{abstract}

\section{Introduction}
Recent studies of gravitational microlensing suggest that
about half of the Galaxy's halo mass can be accounted for
by massive compact halo objects (MACHOs)
with mass $\sim 0.5 M_{\odot}$\footnote{The absence of
short timescale microlensing events
restricts the MACHOs to masses $\gtrsim 0.1 M_{\odot}$,
unless thick disk/light halo models of the Milky Way are
employed. This eliminates brown dwarfs unless, in conjunction
with these nonstandard Galactic models, the brown dwarf mass
function peaks close to the hydrogen burning limit
\cite{alco97}. It has also been suggested that the gravitational
lenses are not in the Galactic halo at all, but rather
in ``the warped and flaring'' Galactic disk \cite{evan97}.} 
\cite{alco97}.
However,
a direct search for halo dwarfs in the Hubble Deep Field
has limited the halo mass fraction of red dwarfs brighter
than magnitude $M_I=15$ to
$f_{\rm RD}<0.06$, and a halo
mass fraction of white dwarfs with $M_I<15$ to
$f_{\rm WD}<0.33$ \cite{flyn96}. Comparing what a present day
white dwarf Galactic halo fraction should have looked like
in the past with deep galaxy surveys leads to a limit of
$f_{\rm WD}<0.10$ \cite{char95}. A study of the infrared
background and Galactic metallicity shows that it is likely
that $f_{\rm WD}<0.25$ \cite{adam96}.

While white dwarf halo mass fractions which exceed these
limits might be obtained with specially tailored assumptions
\cite{adam96,chab96,fiel97}, ultimately it may be necessary to  
consider
more exotic candidates for the MACHOs.
In particular, Jedamzik has recently proposed that horizon-size
black hole formation from primordial density fluctuations
could operate particularly efficiently during a first-order
QCD phase transition \cite{jeda97}. A first-order
confinement transition involves the coexistence of quark-gluon
and hadron phases \cite{qcd}. On large scales,
the pressure response of this
``mixed'' phase is greatly reduced: when compressed, the
energy density does not increase from
increases in pressure and temperature, which remain constant.
Instead, low density hadron phase
is exchanged for high energy density quark-gluon phase.
Because of the reduced pressure response,
the required overdensity for
black hole formation is smaller. For
fluctuations entering the horizon during this epoch,
statistically more abundant, lower amplitude perturbations
could form black holes, leading to a peak in the primordial
black hole (PBH) mass spectrum at the QCD epoch horizon
mass. Since the horizon mass is
\begin{equation}
M_{h}(T)\approx 0.87 M_{\odot} \left(T\over 100{\rm MeV}
	\right)^{-2}\left(g_* \over 51.25\right)^{-1/2},
\label{horizon}
\end{equation}
where $g_*$ is the effective number of relativistic degrees
of freedom at cosmological temperature $T$,
PBHs formed at the QCD transition temperature $T_{\rm QCD}\approx
100 $ MeV could be in the right mass range to be MACHOs.
This possibility takes on added interest in view of
the claim
that galactic halos filled with $\sim 0.5 M_{\odot}$
black holes could bring about enough nearby BH-BH mergers
to cause a few events/year in the first generation network
of gravitational wave interferometers \cite{naka97}.

Other than a general guess that the critical overdensity
for collapse would be reduced by a factor of order unity
during the QCD transition \cite{jeda97,bull97}, no quantitative
estimate of this reduction factor has appeared.
A relativistic numerical calculation is in preparation
\cite{wils98}.
The
purpose of this paper is to use a simplified model to
estimate this reduction factor, and to assess the
cosmological significance of the resulting peak in the
PBH mass spectrum.

\section{Conditions for PBH formation}

\subsection{Radiation equation of state}
A heuristic criterion for gravitational collapse to a black hole
of an overdense region of
energy density $\rho$ and size $S$
was given by Carr and Hawking \cite{carr74}.
These authors argued that collapse would ensue
when the ``gravitational energy'' $\sim G \rho^2 S^5$
was large enough to overcome the ``kinetic energy'' of
expansion and the pressure forces (here $G$ is the gravitational
constant). The overdense
region expands more slowly than the average cosmological
expansion, and eventually turns around and begins
to collapse. Carr and Hawking consider the region at
the moment of turnaround (when the kinetic energy of
expansion is zero), and account for the pressure forces
by comparing the gravitational energy to the ``internal
energy.'' This ``internal energy'' is approximated as
\begin{equation}
p S^3 \sim w \rho S^3, \label{inteng}
\end{equation}
where $p$ is the
pressure and an equation of state of the form $p=w \rho$
is assumed ($w$ is a constant).
Then the condition for an overdense region
to overcome pressure forces and collapse to a black hole
is $G \rho_c S_c^2 \gtrsim w$, where the subscript $c$ indicates
that the quantity is to be evaluated at the moment of
turnaround when the collapse \lq\lq begins.\rq\rq\  This condition
can also be written $S_c \gtrsim (w/G\rho_c)^{1/2}\sim
R_J$, where $R_J$ is the Jeans length. However, the overdense
region cannot be larger than the horizon size at turnaround,
or it will constitute a separate universe, causally
disconnected from our own, and would not be relevant to this
discussion.

For radiation
($w=1/3$), the size of the overdense region at turnaround,
its Schwarzschild radius, the Jeans length,
and the cosmological horizon
size are all of the same order of magnitude.
Therefore it is expected that an overdense region of radiation
satisfying $G \rho_c S_c^2 \gtrsim 1/3$
 will form a black hole of order the horizon mass,
with little chance for centrifugal or turbulent forces to
have a significant effect. For dust (w=0), the Jeans length
is much smaller than the horizon size, and centrifugal
and turbulent forces could prevent full collapse to a black
hole. If a hole does form in this case it could be much
smaller than the horizon mass.

The condition for an overdense region to collapse to a
black hole is expressed above in terms of the density and
size of the region at turnaround. However, cosmological
density perturbations are typically classified by specifying
their overdensity $\delta = (\rho-\bar\rho) /\bar\rho$ when
they enter the horizon, where $\rho$ is the energy
density of the
overdense region and $\bar\rho$ is the average cosmological
density. For assessing the cosmological
significance of the black hole population, it
will therefore be convenient to express the condition for
collapse in terms of $\delta$, which in this paper
will always refer to
the overdensity at horizon crossing.

To make this connection we follow Carr \cite{carr75}.
The line element for the \lq\lq average\rq\rq\  
universe can be taken
to be the conformally flat Friedmann-Robertson-Walker (FRW) form,
\begin{equation}
ds^2=-dt^2+R^2[dr^2 + r^2(d\theta^2 + \sin^2\theta d\phi^2)],
\end{equation}
where the evolution of the time-dependent cosmic
scale factor $R$ is given by
\begin{equation}
\left(dR\over dt\right)^2 ={8\pi G\over 3} \bar\rho R^2.
	\label{friedr}
\end{equation}
Here the \lq\lq average\rq\rq\  background geometry/evolution is
obtained by viewing the universe on a large enough scale that the  
(nearly
critical)
mass-energy density can be
approximated as homogeneously and isotropically
distributed.

We consider ``top-hat'' fluctuations: spherical,
homogeneous overdense regions.
Such overdense regions, if taken to have homogeneously
distributed internal mass-energy, can be described by the metric of
a closed FRW universe,
\begin{equation}
ds^2=-d\tau^2+S^2[{(1-\kappa r^2)}^{-1}dr^2+
	r^2(d\theta^2 + \sin^2\theta d\phi^2)],
\end{equation}
with the evolution of $S$ given by
\begin{equation}
\left(dS\over d\tau\right)^2 ={8\pi G\over 3} \rho S^2
	-\kappa,
\end{equation}
where $\rho$ is the energy density in the overdense region.
Obviously, at some point in the gravitational collapse of such
a fluctuation, pressure gradients will build up and it will not be
possible to approximate the internal
mass-energy distribution as homogeneous. 
However, this approximation
will not be bad until near turnaround. Here $S$ 
can be interpreted as
a sort of \lq\lq radius\rq\rq\ of the evolving fluctuation.
Likewise, $\tau$ is a timelike coordinate characterizing
the evolution of the overdense region.

The general relativistic gauge freedom
inherent in this model is fixed by setting $\tau_h=t_h$,
$S_h=R_h$, and $(dS/d\tau)_h=(dR/dt)_h$, ensuring that
the when the overdense region enters the horizon it
is only a density perturbation.  (The subscript $h$
denotes a quantity evaluated when the overdense region
under consideration enters the horizon.)
At horizon crossing,
the energy density of the fluctuation is $\rho_h=
\bar\rho_h(1+\delta)$. Using these conditions, and employing
an equation of state $p=w\rho$, the evolution of the
overdense region is given by
\begin{equation}
\left(dS\over d\tau\right)^2 ={8\pi G\over 3} \bar\rho_h
	R_h^{3(1+w)}\left[{1+\delta\over S^{1+3w}}-
	{\delta\over R_h^{1+3w}}\right],
\end{equation}
so at turnaround
\begin{equation}
S_c^{1+3w}=R_h^{1+3w}\left(1+\delta\over \delta\right).
\label{sc}
\end{equation}
The quantity appearing in the collapse condition is
\begin{equation}
G \rho_c S_c^2 = {G \bar\rho_h(1+\delta)R_h^{3(1+w)}\over
	S_c^{1+3w}}=G \bar\rho_h R_h^2 \delta,
\end{equation}
where we have used Eq. (\ref{sc}).
Using Eq. (\ref{friedr}) and the fact that the horizon
size $\sim t$,
\begin{equation}
G \bar\rho_h R_h^2 \sim \left(dR\over dt\right)_h^2\sim
\left(R_h\over t_h\right)^2 \sim 1,
\end{equation}
so that the condition for collapse is
\begin{equation}
\delta \gtrsim w,
\end{equation}
or $\delta \gtrsim 1/3$ for radiation. In other words,
in this conventional picture, the minimum fluctuation
amplitude $\delta \rho/\rho$ at horizon crossing required
for collapse to a black hole is of order unity, {\it i.e.} quite  
large.
This large required initial amplitude simply reflects the \lq\lq  
hard\rq\rq\
equation of state of radiation.

\subsection{First-order QCD transition}

It is not at all clear at this point
whether or not there is a first order
cosmic vacuum phase transition
associated with the QCD epoch. The chiral symmetry breaking
event remains the prime candidate for such a 
transition, though the  
lattice
gauge
calculation results on this issue remain unsettled.
For illustrative purposes, we will assume 
that there is a first order  
phase
transition
associated with this epoch in the early universe.
The QCD era is promising for primordial black hole
production, since phase separation at this time implies that the  
duration of a
first order phase transition will be (essentially) a gravitational  
timescale.

A bag-like picture employing a QCD vacuum energy
provides a simple description of a first
order QCD phase transition with critical transition temperature  
$T_{\rm QCD}$.
The pressure and energy density of the hadron phase 
($T \le T_{\rm QCD}$)
are approximately
\begin{equation}
p_{\rm had}(T)={1\over 3} \rho_{\rm had},
\ \ \ \rho_{\rm had}(T)={\pi^2 \over 30}
	g_{\rm had} T^4,
\end{equation}
with, for example, $g_{\rm had} \approx 17.25$.
The quark-gluon phase ($T \ge T_{\rm QCD}$)
is described by
\begin{equation}
p_{\rm qg}(T)={1\over 3}{\pi^2\over 30}g_{\rm qg}T^4 -B,\ \ \
	\rho_{\rm qg}={\pi^2\over 30}g_{\rm qg}T^4 +B,
\end{equation}
where $g_{\rm qg}\approx 51.25$ and $B$ is the QCD vacuum energy  
density.
During the transition ($T\approx T_{\rm QCD}$), the requirement  
$p_{\rm
had}(T_{\rm QCD})
=p_{\rm qg}(T_{\rm QCD})$ yields
\begin{equation}
B={1\over 3}{\pi^2\over 30}(g_{\rm qg}-g_{\rm had})T_{\rm  
QCD}^4.\label{bag}
\end{equation}
Averaged over the coexisting phases,
the energy density during the transition is
\begin{equation}
\langle\rho\rangle=f_{\rm qg} \rho_{\rm qg}(T_{\rm QCD})+
(1-f_{\rm qg})\rho_{\rm had}(T_{\rm QCD}),
\end{equation}
where $f_{\rm qg}$ is the volume fraction in quark-gluon phase.
The evolution of the average energy density during the
transition as a function of cosmic scale factor $R$
can be described as ``dust-like''
($\rho\sim R^{-3}$) \cite{jeda97},
\begin{equation}
\langle\rho\rangle(R)=\left(R_1\over R\right)^3\left(
	\rho_{\rm qg}+{1\over 3}\rho_{\rm had}\right)-{1\over 3}
	\rho_{\rm had},
\end{equation}
where $R_1$ is the scale factor when the cosmic energy
density reaches $\rho_1 \equiv
\rho_{\rm qg}(T_{\rm QCD})$. This is the start of the phase  
transition.

For the purpose of assessing the probability of
black hole formation during the QCD epoch, we will consider
the following further caricature of the behavior of
the plasma through the transition.
For energy densities greater than $\rho_1$,
take the plasma to be radiation with
$g_*=g_{\rm qg}$. This ignores the vacuum energy density $B$, which
at $T=T_{\rm QCD}$ contributes about 18\% of the total. For energy
densities in the range $\rho_2\equiv \rho_{\rm had}(T_{\rm QCD}) <  
\rho
<\rho_1$, we can treat the plasma as dust.
Here $\rho_1$ and $\rho_2$ correspond to the
energy densities in the plasma at the start
and conclusion of the phase transition,
respectively. Even though the pressure
itself is not zero, the almost negligible pressure response
of the plasma during the transition leads to the dust-like
evolution noted above. For densities less than $\rho_2$,
we treat the plasma as radiation with $g_*=g_{\rm had}$.

The duration of the transition can be obtained by comparing
the cosmic scale factor at the beginning ($R_1$) and
end ($R_2$) of the transtion. From the conservation of
co-moving entropy density we obtain
$(R_2/R_1)=(g_{\rm qg}/g_{\rm had})^{1/3}\approx 1.44$, so
the transition is fairly short. An overdense region that
goes through the transition during its collapse does not
correspond to the cases discussed in the previous subsection,
i.e. either radiation or dust throughout the collapse.
How should the condition for collapse to a black hole
 be modified in this
case? Based on the above caricature of the behavior of the
plasma around the transition, we suggest modifying the
``internal energy'' of Eq. (\ref{inteng})
to $(1/3)\rho_c S_c^3 (1-f)$, where
$f$ is the fraction of the overdense region's ``evolution
volume'' spent in the dust-like transition epoch.

For example,
letting $S_1$ and
$S_2$ respectively denote the size of the initially
expanding overdense region  at the beginning and end of the
phase transition, $f=(S_2^3 -S_1^3)/S_c^3$ for an overdense
region that has completed the transition before turnaround.
Put another way, for an overdense region that
obeys $p=w\rho$ for much of its evolution and a dust-like
phase for the rest, we define an effective Jeans length
$R_{J,\rm{eff}}=(w/G \rho_c)^{1/2}(1-f)^{1/2}$ at turnaround.
In terms of the overdensity of the fluctuation at
horizon crossing, the condition for collapse can be expressed
\begin{equation}
\delta \gtrsim w(1-f).\label{critf}
\end{equation}
Therefore, the larger the fraction of a 
fluctuation's evolution time
which is spent in the mixed phase regime,
the smaller the required initial horizon crossing perturbation  
amplitude
for collapse to a black hole.The evaluation of this
collapse criterion will be the subject
of Sec. \ref{sec:crit}. Note that in our analysis we have
assumed that at no time is the universe ever dominated
by the QCD vacuum energy $B$. Were such a
vacuum-dominated condition to arise, then a kind of \lq\lq
mini-inflation\rq\rq\
occurs, where the scale factor can increase exponentially with
time\cite{afm87}.
However, in this case the nucleation rate for bubbles of low  
temperature
phase will increase very rapidly with time and
decreasing temperature, with the result that the fraction $f$ will  
likely
be smaller than in the non-vacuum-dominated case.

\section{Classifying perturbations in the QCD transition epoch}
``Top-hat'' perturbations are completely
specified by $\delta$---their overdensity
when they enter the horizon---and by the cosmological
time at which horizon
crossing occurs. Here we will find it convenient to use
the average cosmological energy density at horizon crossing
$\bar\rho_h$
rather than the time of horizon crossing. In particular,
we use the variable
$x$ to identify the epoch of horizon crossing,
where $x\equiv \bar\rho_h/\rho_1$. Again, $\rho_1$ is
the energy density above which the plasma is pure quark-gluon
phase, and $\rho_2$ is the density below which the plasma
is pure hadron phase. We define the constant $\beta\equiv
\rho_1/\rho_2$. Then perturbations with $x>1$, $\beta^{-1}<x
<1$, and $x<\beta^{-1}$  enter the horizon when
the average cosmological background---{\em not} the overdense
region itself---is in quark-gluon, mixed quark-gluon/hadron,
and hadron phase, respectively.

In addition, our
description of the collapse of an overdense region
involves two important events: (1)
the region's horizon crossing,
and (2) its turnaround, when it stops expanding and begins
to recollapse. The state of the matter at these two events
constitute various classes of perturbations; these are listed
in Table \ref{tab1}.

\begin{figure}
\epsfxsize=5in \epsfbox{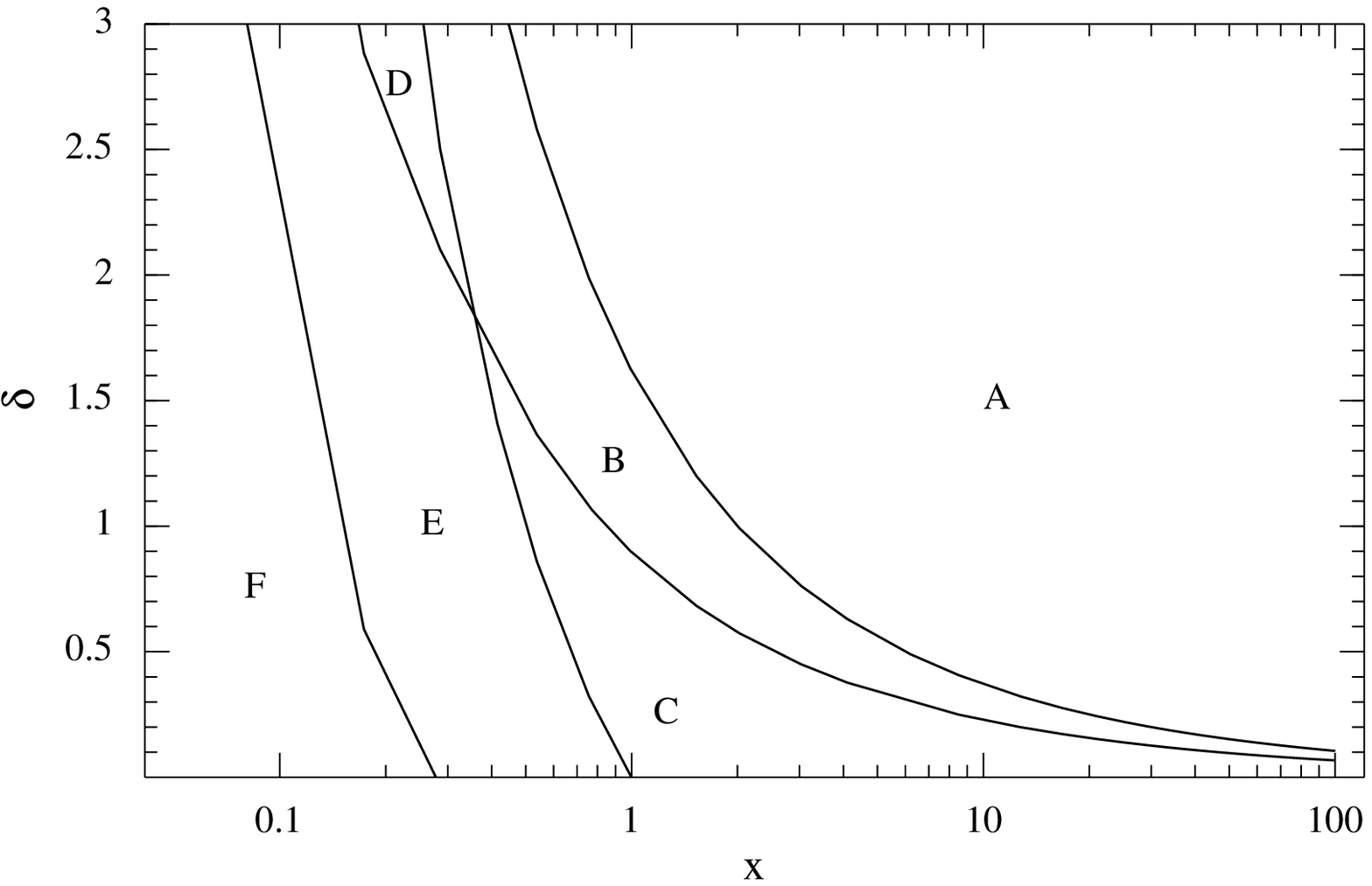}
\caption{Regions in the $(x,\delta)$ plane corresponding
to the classes of perturbations listed in Table \ref{tab1}.
The variable $x$ identifies the epoch a perturbation enters
the horizon:
$x>1\ (x<1)$ corresponds to overdense regions that enter the
horizon before (after) the average cosmological density
begins the transition from quark-gluon to hadron phase.
The quantity $\delta$ is the overdensity $\delta \rho/\bar\rho$
of the perturbation at horizon crossing.}
\label{fig1}
\end{figure}

Given the characteristics of an overdense region,
$\delta$ and $x$, we can specify the class A-F
to which it belongs.
The classes of perturbations that exist for
various epochs of horizon crossing and values of
$\delta$ are shown in Table \ref{tab2}.
 From the information in this table two curves
can be drawn in the $(x,\delta)$ plane (Fig. \ref{fig1}).
Classes A, B, and C are separated from classes D and E
by the curve $\delta=x^{-1}-1$, and classes D and E
are separated from class F by the curve $\delta=\beta^{-1}
x^{-1}-1$. These divisions are based on the state of matter
in the overdense region when it enters the horizon.

In contrast, the curves in the $(x, \delta)$ plane
separating classes A and B, B and C, and D and E
are based on the state of matter in the overdense region
at turnaround. For example, the boundary between classes
A and B is determined by $\rho_{c}=\rho_1$,
that is, by the condition that an overdense region entering
the horizon in the quark-gluon phase just reaches the density
for the phase transition to begin when it begins to recollapse.
The condition
\begin{equation}
\rho_{c}={\bar\rho_{h}(1+\delta) R_h^4 \over S_c^4}=\rho_1,
\end{equation}
using Eq. (\ref{sc}) with $w=1/3$ yields the curve defined by
\begin{equation}
{\delta^2 \over (1+\delta)} =x^{-1}.
\end{equation}
The other curves can be obtained in a similar manner, with
the complication that changes in the density evolution
between radiation ($\rho\propto S^{-4}$) and dust ($\rho
\propto S^{-3}$), and/or vice-versa, must be followed when
the density reaches $\rho_1$ and/or $\rho_2$ respectively.
The analysis leading to Eq. (\ref{sc}) can then be generalized
in a straightforward manner. The curve defining the boundary
between classes B and C is
\begin{equation}
{\delta^3 \over (1+\delta)^{3/2}}=\beta^{-1} x^{-3/2},
\end{equation}
and the boundary between classes D and E is defined by
\begin{equation}
{\delta^3 \over (1+\delta)^2}=\beta^{-1}x^{-1}.
\end{equation}
These last two curves are valid only in the regions
above and below the curve separating classes A, B, and C from
classes D and E, respectively.

Fig. \ref{fig1} shows the
$(x,\delta)$ plane divided into regions that constitute
classes A-F. Class A has no dust-like phase, and $\delta>1/3$
for collapse. Class D does not exist for $\delta \lesssim 1.8$,
so perturbations of this class would be expected to collapse
even without the dust-like phase. Therefore, interesting reductions
in the critical value of $\delta$ for collapse to a black
hole are possible for classes B, C, E, and F.

For reference, the expressions for the turnaround
radius $S_c$ for the various classes are given in Table
\ref{tab3}.
We also give expressions for $S_1$ and
$S_2$, the values of $S$ for
which the density in the overdense region reaches $\rho_1$
and $\rho_2$ respectively. These are
\begin{equation}
S_1=R_h x^{1/4} (1+\delta)^{1/4}, \label{s1}
\end{equation}
which is valid (and useful) for classes B and C,
\begin{equation}
S_2=R_h (\beta x)^{1/3} (1+\delta)^{1/3}, \label{s2}
\end{equation}
valid (and useful) for class E, and 
\begin{equation}
S_2=R_h (\beta x)^{1/4} (1+\delta)^{1/4}, \label{s3}
\end{equation}
for class F.

\section{Computing critical overdensities}\label{sec:crit}
In the framework of our model, overdense regions with
\begin{equation}
\delta\gtrsim {1\over 3}(1-f)
\end{equation}
will collapse to a black hole, where $f$ is the fraction
of  \lq\lq evolution volume\rq\rq\ that the 
overdense region spends  
in the
dust-like phase. For class B, turnaround occurs while
the overdense region is still in the mixed phase:
\begin{equation}
f={S_c^3-S_1^3\over S_c^3}=1-\left({S_1\over S_c}\right)^3
	\ \ \ \rm{(Class\ B)},
\end{equation}
and $S_1$ and $S_c$ are given in Eq. (\ref{s1}) and Table \ref{tab3}
respectively, yielding
\begin{equation}
\left({S_1\over S_c}\right)^3={x^{3/2} \delta^3 \over
	(1+\delta)^{3/2} }.
\end{equation}
For classes C, E, and F, the transition to hadron phase is
completed by turnaround, so
\begin{equation}
f={S_2^3-S_1^3\over S_c^3}\ \ \ \rm{(Classes\ C,\ E,\  
F)}.\label{fcef}
\end{equation}
For class C, it is convenient to write this in the form
\begin{equation}
f=\left(S_1\over S_c\right)^3\left[\left(S_2\over S_1\right)^3
	-1\right]\ \ \ \rm{(Class\ C)},
\end{equation}
where $(S_2/S_1)^3=\rho_1/\rho_2=\beta$, and
\begin{equation}
\left({S_1\over S_c}\right)^3={x^{3/4} \delta^{3/2} \over
	\beta^{1/2} (1+\delta)^{3/4} }\ \ \ \rm{(Class\ C)}.
\end{equation}
For classes E and F it is convenient to write Eq. (\ref{fcef})
in the form
\begin{equation}
f=\left(S_2\over S_c\right)^3\left[1-\left(S_1\over S_2\right)^3
	\right]\ \ \ \rm{(Classes\ E,\ F)},
\end{equation}
where $(S_1/S_2)^3=\beta^{-1}$ and
\begin{eqnarray}
\left({S_2\over S_c}\right)^3={(\beta x)^{1/2} \delta^{3/2} \over
	(1+\delta) }\ \ \ \rm{(Class\ E)} \\
\left({S_2\over S_c}\right)^3={(\beta x)^{3/4} \delta^{3/2} \over
	(1+\delta)^{3/4} }\ \ \ \rm{(Class\ F)},
\end{eqnarray}
using Eqs. (\ref{s2},\ref{s3}) and Table \ref{tab3}.

\begin{figure}
\epsfxsize=5in \epsfbox{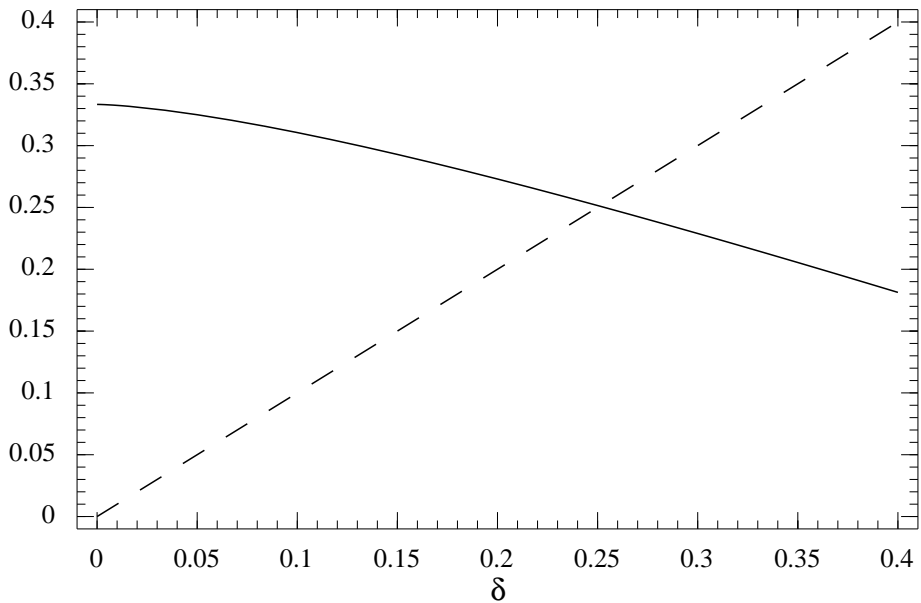}
\caption{Solid curve: $(1-f)/3$, for $x=2$. Dashed curve:
$\delta$. Collapse to a black hole occurs for values of
$\delta$ for which the dashed line is above the solid
curve. The intersection corresponds to the value of $\delta$
at $x=2$ in the curve of Fig. \ref{fig4}.}
\label{fig2}
\end{figure}

\begin{figure}
\epsfxsize=5in \epsfbox{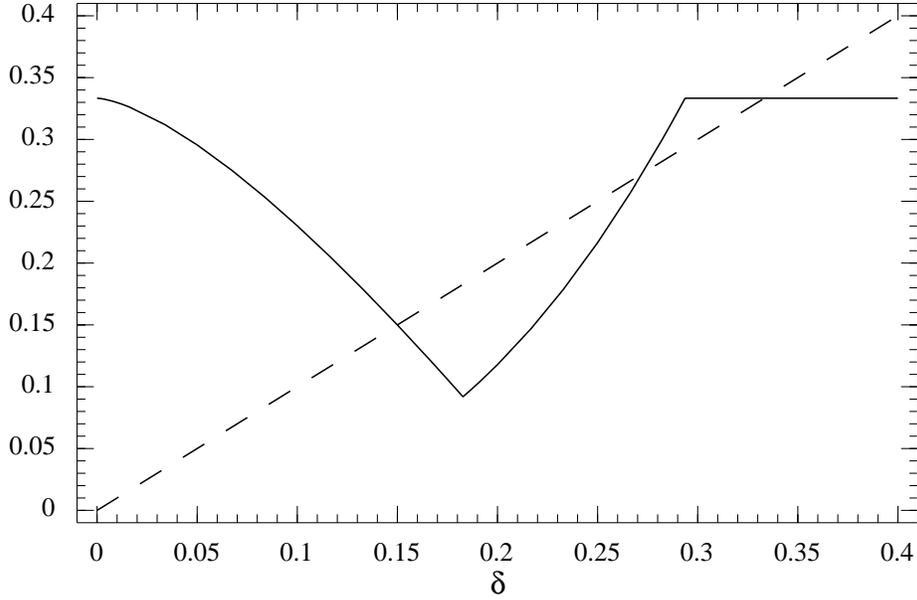}
\caption{Same as Fig. \ref{fig2}, but for $x=15$. The three
intersection points correspond to the values of $\delta$ at
$x=15$ in the curve of Fig. \ref{fig4}.}
\label{fig3}
\end{figure}

For a given $x$---that is, for overdense regions entering the
horizon at a given time---the values of $\delta$
for which collapse
to a black hole will occur can be found by
plotting both $(1-f)/3$ and
$\delta$ itself as functions of $\delta$, taking care to use the
expression for $f$ appropriate to the class
specified by $(x,\delta)$
(see Fig. \ref{fig1}). Figure \ref{fig2} shows such a plot for
$x=2$, from which it is clear that collapse will occur for all
values of $\delta$ above a certain critical value. However,
as seen in Fig. \ref{fig3}, for larger
values of $x$ ($x=15$ in Fig. \ref{fig3}) an interesting thing  
occurs:
the range of $\delta$ for which $\delta>(1-f)/3$
has an upper as well
as lower bound. Figure \ref{fig4} indicates the region in the
$(x,\delta)$ plane for which collapse to a black hole occurs. The
range of $x$ for which there is a reduction in the
critical overdensity for collapse has an upper
bound; and while it
has no lower bound, the reduction becomes
slight at small values of $x$.

\section{Cosmological significance}
The potential cosmological significance of primordial
black holes lies in the fact that the energy density
of a black hole population redshifts as $R^{-3}$, in
contrast with radiation, whose energy density redshifts
as $R^{-4}$. Thus the relative contribution of black holes
to the cosmological density parameter $\Omega$ increases
as the universe expands. See, for example,
the work by Crawford and Schramm\cite{cs82}.

For black holes formed
at a particular epoch, the contribution to the present-day
density parameter is
\begin{equation}
\Omega_{\rm{BH}}\approx 2\times 10^7 \epsilon(T) \left({T_0
	\over 2.75 \rm{K}}\right)^3\left({T\over 100
	\rm{MeV}}\right) h^{-2}, \label{omega}
\end{equation}
where $\epsilon(T)$ is the fraction of radiation energy density
converted into black holes at cosmological temperature
$T$, $T_0$ is the present-day microwave background temperature,
and $h$ is the present epoch Hubble parameter in units of 100 km
s$^{-1}$ Mpc$^{-1}$. Assuming that the mass of the black
holes is the horizon mass at $T$, we can write
\begin{equation}
\epsilon(T)=\int_{\delta_a}^{\delta_b} F(\delta, T)\,
	d\delta, \label{eps}
\end{equation}
where $F(\delta,T)$ is the probability of a
horizon volume at $T$ to have
overdensity $\delta$, and $\delta_a$ and $\delta_b$
delimit the range of overdensities that contribute to the
black hole population.

The power spectrum (spatial Fourier transform) of
density perturbations is typically assumed to have a
power law form, $|\delta_k|^2\propto k^n$, where $k$ is
the wavenumber and $n$ is the spectral index. Density
perturbations generated during inflation typically have
$n\approx 1$, with a normal distribution of overdensities:
\begin{equation}
F={1\over \sqrt{2\pi}\sigma(M)} \exp\left(-{\delta^2
\over 2 \sigma^2(M)}\right),
\end{equation}
where $\sigma(M)$ is the variance at horizon mass $M$ and
can be taken to be \cite{gree97}
\begin{equation}
\sigma(M)=9.5\times 10^{-5}\left({M\over 10^{22}M_{\odot}}
\label{sigma}
\right)^{(1-n)/4}
\end{equation}
when normalized by COBE measurements of the microwave
background anisotropy.

\begin{figure}
\epsfxsize=5in \epsfbox{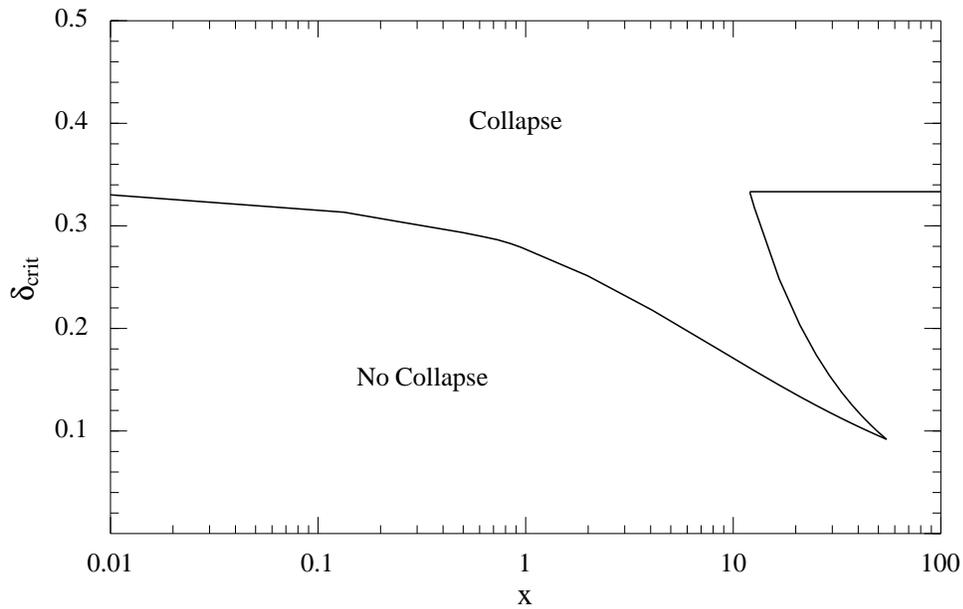}
\caption{The curve in the $(x,\delta)$ plane indicating which 
	parameter values lead to collapse to a black hole.
	Without the phase transition, this would be a 
	straight line at $\delta=1/3$.}
\label{fig4}
\end{figure}

The critical value of $\delta$ above which collapse to a
black hole can occur [$\delta_a$ in Eq. (\ref{eps})] is given
by the curve in Fig. \ref{fig4}, taking the lowest branch when
it becomes multivalued at the higher values of $x$. Noting that
perturbations with $\delta\gtrsim 1/3$ make negligible
contributions to $\Omega$, we take $\delta_b=1/3$, except for
the higher values of $x$ for which the the curve in Fig.
(\ref{fig4}) is multivalued. For this region, $\delta_b$ is
given by the middle branch of the curve in Fig. (\ref{fig4}).

It is convenient to change from our variable
$x=\bar\rho_h/\rho_1$,
in terms of which the curve in Fig. (\ref{fig4}) was calculated,
to the horizon mass $M$.
This connection is made as follows. We account for the fact
that we have ignored the vacuum energy density $B$ in the
quark-gluon phase by defining an effective transition temperature
$T_{\rm QCD,eff}$ by
\begin{equation}
{\pi^2\over 30}g_{\rm qg} T_{\rm QCD}^4 + B = 
{\pi^2\over 30}g_{\rm  
qg}
	T_{\rm QCD,eff}^4.
\end{equation}
Using this equation and Eqs. (\ref{horizon}) and (\ref{bag}),
we then have
the following relation between $x$ and $M$:
\begin{equation}
x={g_* T^4\over g_{\rm qg} T_{\rm QCD,eff} }=\left({
	{4\over 3}g_{\rm qg}-{1\over 3} g_{\rm had} \over
	51.25}\right)^{-1} \left({M\over 0.87 M_{\odot}}
	\right)^{-2}\left({T_{\rm QCD}\over 100  
\rm{MeV}}\right)^{-4}.
\end{equation}
For a given value of the spectral index $n$, $\epsilon(M)$
can be computed, where the horizon mass
$M$ has replaced $T$ in Eq.
(\ref{eps}) as the variable indicating the cosmological epoch.

For the values of $n$ we will have to consider ($n\approx 1.4$),
$\epsilon(M)$ is a very strongly peaked function,
justifying the assumption in Eq. (\ref{omega}) of black holes
being formed at a single epoch. This sharp peak comes about as
follows. For $n>1$, there is increasing power on small scales
[Eq. (\ref{sigma})]. Also, $\delta_a$ decreases with decreasing
$M$ (increasing $x$, see Fig. \ref{fig4}). Both of these
tend to make $\epsilon(M)$ increase rapidly with decreasing $M$.
However, when $M$ gets too small, the range of
overdensities contributing to black holes is cut off completely
[the cusp at $x\approx 50$ in Fig. (\ref{fig4})], and
$\epsilon(M)$ drops abruptly.

To obtain $\epsilon$ for use in Eq. (\ref{omega}),
we  choose a value of $n$ and find the peak value of
$\epsilon(M)$. The mass of the black holes is the value
of $M$ for which $\epsilon$ has its peak value. Figures
(\ref{fig5}) and (\ref{fig6}) show $\Omega_{\rm{BH}}$
as a function of $n$ for two choices of
transition temperature $T_{\rm QCD}$. The fine tuning in $n$
required for $\Omega_{\rm{BH}}\sim 1$ is immediately apparent
\cite{bull97}.
For a given $T_{\rm QCD}$
the masses of the black holes do not vary significantly
over the ranges of $n$ displayed:
$m_{\rm{BH}}\simeq 0.11 M_{\odot}$ for $T_{\rm QCD}=100$ MeV,
and $m_{\rm{BH}}\simeq 0.03 M_{\odot}$ for $T_{\rm QCD}=200$ MeV.

\section{Conclusions}

Given the potential difficulties in accounting for
the observed MACHO population with red or white dwarfs,
it is interesting to follow up on the suggestion \cite{jeda97}
that the MACHOs could be black holes formed during a
first-order QCD confinement transition. We have performed
a semianalytic analysis using a simplified model. This
model involves consideration of spherical ``top-hat'' overdense
regions and the use of the Friedmann equation to calculate
their evolution. A caricature of the behavior of the equation
of state near the QCD transition is employed, along with
an
extension of the condition given by Carr and Hawking
\cite{carr74} for primordial overdense regions
to collapse to black holes.

\begin{figure}
\epsfxsize=5in \epsfbox{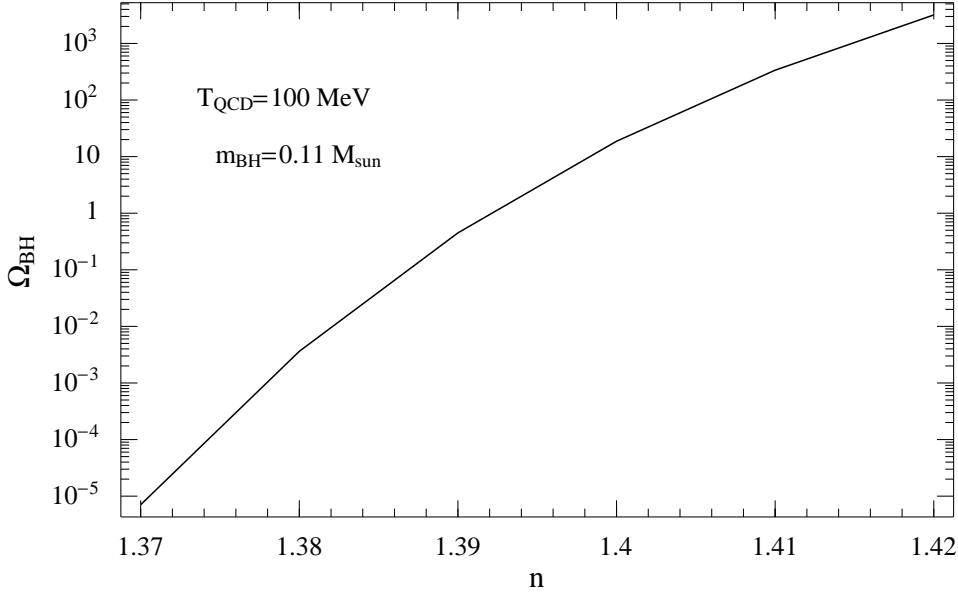}
\caption{The contribution to the present day density parameter
	by black holes formed during the QCD transition, as
	function of the spectral index of the primordial
	density fluctuations.
	The transition temperature is taken to be
	$T_{\rm QCD}=100$ MeV, and the resulting black holes have
	mass $\approx 0.11 M_{\odot}$. }
\label{fig5}
\end{figure}

\begin{figure}
\epsfxsize=5in \epsfbox{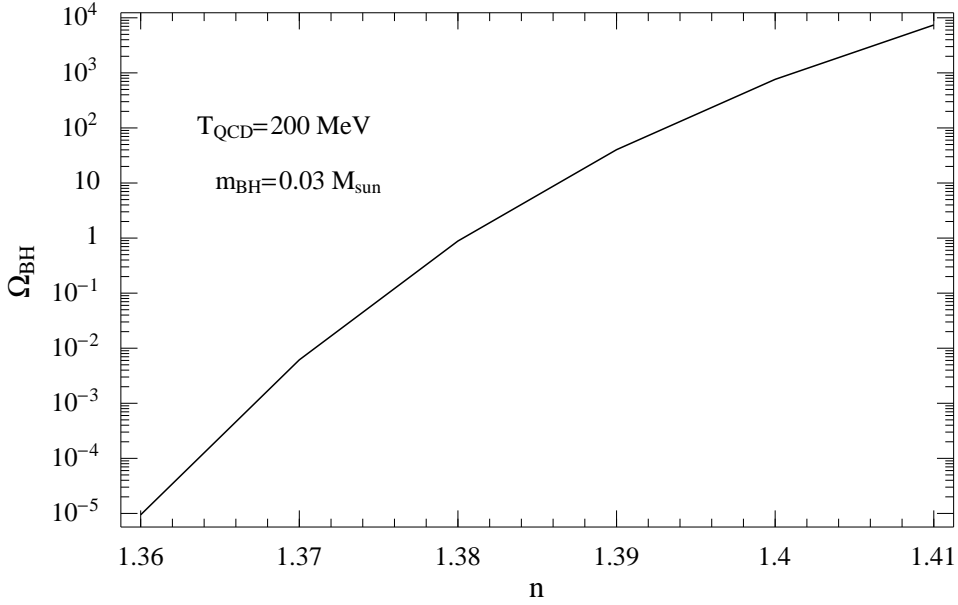}
\caption{Same as Fig. \ref{fig5}, but for
	transition temperature 
	$T_{\rm QCD}=200$ MeV. The black holes in this case have
	mass $\approx 0.03 M_{\odot}$. }
\label{fig6}
\end{figure}

In the context of a COBE normalized primordial density
perturbation spectrum characterized by a power law
with constant spectral index $n$, we find that for the
black holes so produced to be cosmologically significant
would require $n \sim 1.4$, with extreme fine tuning
(see Figs. \ref{fig5}-\ref{fig6}) \cite{jeda97,bull97}.
While such a spectral
index is not in conflict with observations of microwave
background anisotropies \cite{benn96,hu94},
it conflicts with limits on $n$
based on, for example, evaporation of lower mass black
holes formed at earlier epochs (e.g., Refs.
\cite{gree97,carr94} and references therein).
In addition, we find that
the holes most efficiently formed by this mechanism
entered the horizon at an epoch somewhat
earlier than the time the average universe goes through
the QCD transition. Accordingly, we find black holes masses
in the range $0.11-0.03 M_{\odot}$ for QCD transition
temperatures of $100-200$ MeV. This mass range is somewhat
low for the estimated $0.5 M_{\odot}$ MACHO mass.

Our model involves some gross simplifications. The condition
for collapse to a black hole is heuristic. The use of
spherical ``top-hat'' perturbations is not realistic to begin
with, and using the Friedmann equation for their
evolution after they enter the horizon is even less realistic.
The bag-like model is not an overly detailed description of the
behavior of matter near the confinement transition, and we have
even employed a caricature of the bag model. We have assumed,
in keeping with the standard lore, that the mass
of black holes formed
during a radiation dominated era will be roughly
the horizon mass at
the epoch the overdense region enters the horizon.

The results from our simplified model are not particularly
encouraging for this mechanism of black hole formation, and
refinements in the treatment would probably make matters worse.
Early numerical treatments of black hole collapse had
different results, with one set of calculations
\cite{bick79} finding black holes to have about the horizon
mass, while othere calculations \cite{nade78} found that
the holes formed with masses somewhat
smaller than the horizon mass.
More recent calculations \cite{niem97}
suggest that, contrary to the usual
assumption, the masses of primordial black holes formed in a
radiation dominated
era may be significantly smaller than the horizon mass. These
same calculations also found that the overdensities
necessary for
collapse in a radiation era
are 2-3 times higher than the $\delta\approx 1/3$
from the standard lore. (Larger values of $\delta$ are
also obtained if one keeps various factors of order unity in
the heuristic analysis leading to the condition
$\delta\gtrsim 1/3$ for collapse.)
Finally, in assessing the cosmological
significance of the black holes we have assumed a Gaussian
distribution of overdensities. Non-gaussian effects on the
large overdensity tail are likely to be skew-negative
\cite{bull97}, although they do not weaken the effect on
the needed value of the spectral index too much \cite{gree97}.

The shortcomings of our model may make it unsuitble for
precise quantitative predictions, but we believe two
conclusions can be drawn from it that would be confirmed with
refined treatment: 1) The perturbations gaining the most
``benefit'' from a first-order QCD transition enter the horizon
somewhat earlier than the time the averaged universe goes through
the transition. (This effect is also being seen in
numerical calculations \cite{jeda97b}.) 
This means that the mass of the black
holes formed by this mechanism will be somewhat smaller than
the horizon mass when the averaged universe goes through the
transition. 2) If density perturbations can be characterized
as a COBE-normalized power law with constant spectral index,
a very finely tuned, blue spectrum would be required for the
black holes produced by this mechanism to be cosmologically
significant \cite{jeda97,bull97}.

\acknowledgements

We thank Karsten Jedamzik for communications regarding 
ongoing calculations. We also thank Mitesh Patel and Jim Wilson
for useful conversations. This work was supported by grant
DOE-FG02-87ER40317 at SUNY at Stony Brook, and by 
NSF PHY95-03384 and NASA NAG5-3062
at UCSD.


%
%

%
%
\begin{table}
\caption{Classification of overdense regions according to the
state of matter at horizon crossing and turnaround.}
\label{tab1}
\begin{tabular}{ccc}
Class & Horizon crossing\tablenotemark[1]
	& Turnaround\tablenotemark[1] \\
\hline
A & qg & qg \\
B & qg & m \\
C & qg & h \\
D & m  & m \\
E & m  & h \\
F & h  & h
\end{tabular}
\tablenotetext[1]{{\em gq:} quark-gluon phase. {\em m:}
	mixed phase, i.e. coexistent quark-gluon and hadron
	phases. {\em h:} hadron phase.}
\end{table}

\begin{table}
\caption{Classes of perturbations that exist for
various epochs of horizon crossing and ranges of
$\delta$.}
\label{tab2}
\begin{tabular}{lll}
Epoch of orizon crossing
	& Overdensity & Class \\
\hline
$x>1$ & $\delta>0$ & A, B, C \\
\hline
$\beta^{-1}<x<1$ & $\delta>x^{-1}-1$ & A, B, C \\
	& $0<\delta<x^{-1}-1$ & D, E \\
\hline
 $x<\beta^{-1}$ & $\delta>x^{-1}-1$ & A, B \\
 & $\beta^{-1}x^{-1}<\delta<x^{-1}-1$ & D, E \\
 & $0<\delta<\beta^{-1}x^{-1}$  & F
\end{tabular}
\end{table}

\begin{table}
\caption{Radius of the overdense region at turnaround.}
\label{tab3}
\begin{tabular}{ll}
Class & Radius at turnaround \\
\hline
A & $S_c=R_h\left({1+\delta\over \delta}\right)^{1/2}$ \\
B & $S_c=R_h x^{-1/4}{(1+\delta)^{3/4} \over \delta }$ \\
C & $S_c=R_h \beta^{1/6}
	\left({1+\delta\over \delta}\right)^{1/2}$   \\
D & $S_c=R_h {(1+\delta)\over \delta}$ \\
E & $S_c=R_h (\beta x)^{1/6} {(1+\delta)^{2/3} \over
	\delta^{1/2}}$ \\
F & $S_c=R_h\left({1+\delta\over \delta}\right)^{1/2}$
\end{tabular}
\end{table}


\begin{references}
\bibitem{alco97}C. Alcock et al. (MACHO Collaboration),
	Astrophys. J. {\bf 486}, 697 (1997).
\bibitem{evan97}N. W. Evans, G. Gyuk, M. S. Turner, and
	J. J. Binney, Report No.
	astro-ph/9711224, 1997 (unpublished).
\bibitem{flyn96}C. Flynn, A. Gould, and J. N.
	Bahcall, Astrophys. J.
	{\bf 466}, L55 (1996).
\bibitem{char95}S. Charlot and J. Silk, Astrophys. J {\bf 445},
	124 (1995).
\bibitem{adam96}F. C. Adams and G. Laughlin,
	Astrophys. J. {\bf 468},
	586 (1996).
\bibitem{chab96}G. Chabrier, L. Segretain, and D. M\'era,
	Astrophys. J. {\bf 468}, L21 (1996).
\bibitem{fiel97}B. D. Fields, G. J. Mathews, and D. N. Schramm,
	Astrophys. J. {\bf 483}, 625 (1997).
\bibitem{jeda97}K. Jedamzik, Phys. Rev. D {\bf 55},
	R5871 (1997).
\bibitem{qcd}E. Witten, Phys. Rev. D {\bf 30}, 272 (1984);
	J. H. Applegate and C. Hogan, Phys. Rev. D {\bf 31},
	3037 (1985); G. M. Fuller, G. J. Mathews, and C. R.
	Alcock, Phys. Rev. D {\bf 37}, 1380 (1988).
\bibitem{naka97}T. Nakamura, M. Sasaki, T. Tanaka, and
	K. S. Thorne, Astrophys. J. {\bf 487}, L139 (1997).
\bibitem{bull97}J. S. Bullock and J. R. Primack, Phys. Rev.
	D {\bf 55}, 7423 (1997).
\bibitem{wils98}J. R. Wilson, G. M. Fuller, and C. Y. Cardall,
	in preparation (1997).
\bibitem{carr74}B. J. Carr and S. W. Hawking, Mon. Not. R. Ast.
	Soc. {\bf 168}, 399 (1974).
\bibitem{carr75}B. J. Carr, Astrophys. J. {\bf 201}, 1 (1975).
\bibitem{afm87}C. R. Alcock, G. M. Fuller, and G. J. Mathews,
          Astrophys. J. {\bf 320}, 439 (1987).
\bibitem{cs82}M. Crawford and D. N. Schramm, Nature {\bf 298}, 538
           (1982).
\bibitem{gree97}A. M. Green and A. R. Liddle, Report No.
	SUSSEX-AST 97/4-2, astro-ph/9704251, 1997 (unpublished).
\bibitem{benn96}C. L. Bennett et al., Astrophys. J. {\bf 464},
	L1 (1996).
\bibitem{hu94}W. Hu, D. Scott, and J. Silk, Astrophys. J. {\bf
	430}, L5 (1994).
\bibitem{carr94}B. J. Carr, J. H. Gilbert, and J. E. Lidsey,
	Phys. Rev. D {\bf 50}, 4853 (1994).
\bibitem{bick79}G. V. Bicknell and R. N. Henriksen,
	Astrophys. J. {\bf 232}, 670 (1979).
\bibitem{nade78}D. K. Nadezhin, I. D. Novikov, and
	A. G. Polnarev, Sov. Astron. {\bf 22}, 129 (1978);
	I. D. Novikov and A. G. Polnarev, in {\em Sources
	of Gravitational Radiation, Proceedings of the
	Battelle Seattle Workshop,} ed. L. L. Smarr
	(Cambridge University Press, Cambridge, 1979), 173.
\bibitem{niem97}J. C. Niemeyer and K. Jedamzik, Report No.
	astro-ph/9709072, 1997 (unpublished).
\bibitem{jeda97b}K. Jedamzik, private communication.
\end{references}
\end{document}